\documentclass{article}
\usepackage{amssymb}
\usepackage{graphicx}

\input{tcilatex}
\begin{document}

\title{An agent based multi-optional model for the diffusion of innovations}
\author{Carlos E. Laciana and Nicol\'{a}s Oteiza-Aguirre \\
\\
Grupo de Aplicaciones de Modelos de Agentes (GAMA),\\
Facultad de Ingenier\'{\i}a, Universidad de Buenos Aires, \\
Avenida Las Heras 2214 Ciudad Aut\'{o}noma de Buenos\\
Aires, C1127AAR, Argentina. \\
clacian@fi.uba.ar 
Decision under uncertainty; Potts model; \\
Heterogeneous population; Innovation diffusion\ \ }
\date{ }
\maketitle

\begin{abstract}
We propose a model for the diffusion of several products competing in a
common market based on the generalization of the Ising model of statistical
mechanics (Potts model). Using an agent based implementation we analyze two
problems: (i) a three options case, i.e. to adopt a product A, a product B,
or non-adoption and (ii) a four option case, i.e. the adoption of product A,
product B, both, or none. In the first case we analyze a launching strategy
for one of the two products, which delays its launching with the objective
of competing with improvements. Market shares reached by each product are
then estimated at market saturation. Finally, simulations are carried out
with varying degrees of social network topology, uncertainty, and population
homogeneity.
\end{abstract}

\section{Introduction}

The process of diffusion of innovations has motivated ample academic
interest in the decade of the sixties, extending until today. A pioneer work
has been the one developed by Bass \cite{Bass 69}. Originally, this model
was fitted, with enough precision, to the observational data of adoption
rates corresponding to many consumer durable goods, obtaining the cumulative
S-curves for the number of adopters, in which fast growth is generated by
word of mouth between early and late adopters \cite{Rogers}. Later, the Bass
model was extended in its use to other products, for example those related
to the telecommunications sector \cite{Bass 2004}. Many of the innovation
diffusion models were inspired by models initially developed for application
in different disciplines. The Bass model was not the exception; in that case
the analogy was made with epidemic models \cite{Geroski}.

The Bass model performs an aggregate description of the behavior of
potential decision makers in relation with the adoption or not of an
innovation (or technology). In this formulation the set of deciders is
assumed to be homogeneous and totally connected, or equivalently, we suppose
that each individual is influenced by all remaining decision makers. The
basic assumption of the model is that at each point in time, potential
buyers are exposed to two kinds of influences: an external influence, in the
shape of advertising campaigns carried out by the companies via the mass
media; and an internal influence coming from "word-of-mouth" interactions
with adopters \cite{Mahajan 1990}.

The Bass model has the general advantage of its simple application,
constituting a description on a macro level that shows the observed global
behavior \cite{Peres}. However, the lack of detail at a microscopic level
turns it into a weak prediction instrument, restricting its application to
analogies with similar known products \cite{Bass 2004}.

On the other hand, the study of diffusion models at micro level has recently
intensified, with special focusing on agent based models (ABMs), as is
shown, for example, in the review paper by \cite{Kiesling}. In general, ABMs
involve two complementary causes for innovation adoption: a) the individual
evaluation of the advantages related to the adoption of the new product (or
technology) and b) the imitation of behavior of close contacts which are
considered as examples to follow by the decision maker. ABMs have the
advantage, compared to macro models, that the introduction of the social
networks is possible. Those social networks are the routes for interaction
between the elements of the system \cite{Kuandykov}.

In the case of the diffusion of only one new product in a potential market,
individuals in ABMs must choose between two alternatives: to adopt or not to
adopt. It was found that the social system described before, can be obtained
by using the analogy with the well-known Ising statistic model \cite%
{Bourgine}, which was originally developed in the field of physics to
describe the phase transition in ferromagnetic materials \cite{Ising}. In
the physical model the agents represent spins in a regular lattice. Those
spins can be in two positions: one in the direction of an external field and
the other in the opposite direction. In that model, each agent (ion in the
metallic lattice) can be influenced by its neighbors, which generate a local
field that induce the orientation of the spin. There are examples in
scientific literature in which the Ising model is adapted to modeling the
social process of opinion formation \cite{Grabowski} \cite{Galam} and others
in which the diffusion of technology is studied \cite{Laciana}.

In the Ising model, the interaction is limited to the nearest neighbors in a
regular array. However, a social process of communication not necessarily
corresponds to an interaction with the geographic proximity, so it is
necessary to modify the formalism in order to include irregular networks
where the notion of nearness is diffuse. Such is the case of the family of
networks known as "small world networks" (SWN) described by Watts and
Strogatz \cite{Watts Strogatz}. The different networks in the family of SWN
are obtained by the variation of only one parameter called rewiring
probability. The idea is to reconnect the agents, starting from a regular
network with a probability which is the parameter mentioned before. A
rewiring probability of zero corresponds to a regular network, and a
probability of one to a totally random network. For values of probability in
the interval [0.003, 0.02], the SWN for a two dimensional network, are
obtained \cite{Xu}. SWN have been strongly studied in the scientific
literature, observing that they have suitable properties for applications in
the social communications and other natural phenomena \cite{Albert}.

At this point we have an Ising-like model equipped with a SWN which lets us
study the technology diffusion process corresponding to only one product, as
was applied in the previous work \cite{Laciana}. However the generalization
of this model to more products, whose physical analogy is the Potts model,
has been used in few cases of the social field \cite{Vega}. The idea of the
present work is the exploration of the possibilities of the Potts-like
formulation for the analysis of the diffusion process associated to the
launching of "M" products in a common market.

The case in which two brands compete for a unique market, is analyzed in 
\emph{\ }ref.\emph{\ }\cite{Libai 2013}\emph{, }by using an agent-based
cascade model. Following refs.\emph{\ }\cite{Kempe}\emph{\ }\cite{Leskovec},
we can identify two mayor categories of agent-based diffusion models:
a)threshold models, in which agents adopt when a specified minimum number of
neighbors have adopted, and b) cascade models, where the probability of
adoption increases with the number of adopters in the neighborhood, with an
exponential mathematical dependency.

One contribution of our proposal, considering the two categories given
before, is that this formalism could be thought of as a threshold model, but
applicable to \textquotedblleft M\textquotedblright\ options, and
particularly applied in this paper to systems of three and four options. In
the algorithm we propose, adapted from statistical mechanics applications,
agents have an implicit threshold that relates to the perceived effective
utility of each of the potential choices. This threshold is a function of
the differences of utility of the available options, as is shown in ref.%
\emph{\ }\cite{Laciana}\emph{. }Moreover, this model can also generate a
more stochastic decision process, slightly moving away from the
\textquotedblleft threshold\textquotedblright\ category, when uncertainty is
considered by varying the temperature parameter.

Yet another contribution of this paper lies in the way agents decide between
the available choices, which is embedded in the model. That means that the
probability of adoption is dynamically modified in time for each agent,
depending of the consumption choices of their particular set of neighbors.
This methodology is different from, for example, the ones used by \cite%
{Libai 2005} \cite{Goldenberg 2001}\emph{, }where diffusion is described as
a two stage process; first, the units adopt, and then, at each point in
time, adopters can decide whether to disadopt, or switch, according to a
separate \textquotedblleft coin-flip\textquotedblright .\emph{\ }

The work is organized as follows: section 2 gives the formalism for "M"
options, showing also how the Ising-like (M=2) and the studied cases (M=3
and M=4) can be deduced from it. Section 3 describes the variables of the
model, such as: rewiring probability, the utility of each product, the
spatial and temporal distribution of early adopters, and the temperature as
a measure of uncertainty in the decision. Section 4 gives the details about
the implementation of the agent based model. Section 5 shows and performs
the analysis of the results obtained. Section 6 summarizes the main
conclusions.

\section{Formalism for "M" options (Potts-like)}

Let us consider a population of N agents, which are identified by Greek
indexes, and M possible states, which are identified by Latin indexes, for
all the agents. The set of states $E$ is written as: $E=\left\{ 
\overrightarrow{X}_{1},...,\overrightarrow{X}_{M}\right\} $ with $%
\overrightarrow{X}_{i}\in R^{M}.$ For simplicity we choose a canonical
basis, then $E=\left\{ \left( 1,0,...,0\right) ;\left( 0,1,0,..,0\right)
;...;\left( 0,0,...,1\right) \right\} .$

Introducing the scalar product between a pair of states, we have

\begin{equation}
\overrightarrow{X}_{i}\cdot \overrightarrow{X}_{j}=\delta _{ij},  \label{1}
\end{equation}

where the term on the r.h.s. is the Kr\"{o}eneker delta.

Because the agent will be in some of the states of the set $E$, it is
convenient to define a time-dependent vectorial function, $\overrightarrow{S}%
_{\alpha }\left( t\right) $, where the discrete variable $\alpha $\
identifies each agent, then, $\alpha =1,2,...,N$, and $N$ is the total
number of agents. For example, if agent $\alpha $ is in the $k$ state, at
time t, we have

\begin{equation}
\overrightarrow{S}_{\alpha }\left( t\right) =\overrightarrow{X}_{k}.
\label{2}
\end{equation}

The interpretation of the state $k$ could be the adoption of a given product
(we can directly call it product $k$), then the dimension of the space
generated by $E$ coincides with the number of options.

Now we introduce the probability to find the agent $\alpha $, at the instant 
$t$ in the state $k$, indicated as $P(\overrightarrow{S}_{\alpha }\left(
t\right) =\overrightarrow{X}_{k})$. We will assume, in order to re-obtain
the Ising model when $N=2$, that the probability to finding agent $\alpha $
in the $k$ state in a given instant, is given by the Boltzmann-Gibbs
distribution, i.e.

\begin{equation}
P\left( \overrightarrow{S}_{\alpha }\left( t\right) =\overrightarrow{X}%
_{k}\right) =\frac{e^{\beta \overrightarrow{m}_{\alpha }\cdot 
\overrightarrow{X}_{k}}}{\tsum\limits_{j=1}^{M}e^{\beta \overrightarrow{m}%
_{\alpha }\cdot \overrightarrow{X}_{j}}},  \label{3}
\end{equation}

where, as in the Ising model, $\beta =1/\kappa T$, $\kappa $ is the
Boltzmann constant and $T$ is\ the temperaure. \ In our calculation we will
consider $\kappa =1.$

We define vector $\overrightarrow{m}_{\alpha }$ by

\begin{equation}
\overrightarrow{m}_{\alpha }\left( t\right) =\frac{1}{V_{\alpha }\left(
t\right) }\sum_{\gamma =1}^{N}J_{\alpha \gamma }\left( t\right) 
\overrightarrow{S}_{\gamma }\left( t\right) +\overrightarrow{u},  \label{4}
\end{equation}

$V_{\alpha }\left( t\right) $ is the number of contacts of agent $\alpha $
at time $t$. The coefficients $J_{\alpha \gamma }\left( t\right) $ allow the
connection between agent $\alpha $ and agent $\gamma .$ Those coefficients,
in the simpler form, are zeros or ones, where 1 indicates the connection
between the respective agents exists. In this case, all the connections are
equally weighted. The distribution of the zeros and ones in the matrix $%
J\equiv \left[ J_{\alpha \gamma }\right] $ describes the social network
topology.

Vector $\overrightarrow{u}$ is defined as the utility vector. It has a
dimension equal to the number of states i.e. M, being each one of its
elements the utility that corresponds to the election of each one of the M
products (or states), so

\begin{equation}
\overrightarrow{u}=\left( u\left( \overrightarrow{X}_{1}\right) ,...,u\left( 
\overrightarrow{X}_{k}\right) ,...,u\left( \overrightarrow{X}_{M}\right)
\right) .  \label{5}
\end{equation}

The scalar product between $\overrightarrow{u}$ and the state vector $%
\overrightarrow{X}_{k}$ shows the utility corresponding to that state.
Mathematically, this can be expressed as

\begin{equation}
\overrightarrow{u}\cdot \overrightarrow{X}_{k}=u\left( \overrightarrow{X}%
_{k}\right) ,  \label{6}
\end{equation}

because is $\overrightarrow{X}_{k}=\left( 0,...,0,1,0,...,0\right) $ with
the k$^{th}$ component equal to one.

Dividing in the numerator and denominator of Eq. (\ref{3}) by $\exp \left(
\beta \overrightarrow{m}_{\alpha }\cdot \overrightarrow{X}_{k}\right) $ the
expression can be re-written in the following way to simplify its analysis:

\begin{equation}
P\left( \overrightarrow{S}_{\alpha }\left( t\right) =\overrightarrow{X}%
_{k}\right) =\frac{1}{1+\tsum\limits_{j\neq k}e^{-\beta \Delta _{kj}}},
\label{7}
\end{equation}

where the exponent has the form

$\Delta _{kj}\equiv \overrightarrow{m}_{\alpha }\cdot \left( \overrightarrow{%
X}_{k}-\overrightarrow{X}_{j}\right) $%
\begin{equation}
=\frac{1}{V_{\alpha }\left( t\right) }\sum_{\gamma =1}^{N}J_{\alpha \gamma
}\left( t\right) \overrightarrow{S}_{\gamma }\left( t\right) \cdot \left( 
\overrightarrow{X}_{k}-\overrightarrow{X}_{j}\right) +u\left( 
\overrightarrow{X}_{k}\right) -u\left( \overrightarrow{X}_{j}\right) ,
\label{8}
\end{equation}

The first term on the r.h.s. of Eq. (\ref{8}) represents the social
contribution due to the imitation effect. If our interest is in the adoption
dynamics of product $k$, we can see that when $\overrightarrow{S}_{\gamma
}\left( t\right) \cdot \overrightarrow{X}_{j}$ $\neq 0$ we obtain a negative
contribution in regards to adoption, which is produced by the existence of
some non-adopters within the group of contacts of the considered agent. The
other term in Eq. (\ref{8}) is the difference between the utilities of
products $k$ and $j$.

In a more intuitive way $\Delta _{kj}$ can be written as

\begin{equation}
\Delta _{kj}=\nu _{k}-\nu _{j}+u\left( \overrightarrow{X}_{k}\right)
-u\left( \overrightarrow{X}_{j}\right) \equiv \Delta \nu _{kj}+\Delta u_{kj},
\label{9}
\end{equation}

where $\nu _{k},$ $\nu _{j}$ \ represent the proportions of adopters of $k$
and $j$ respectively within the set of contacts of the agent considered
(nearest neighbors in the particular case of the regular lattice). The first
subtraction on the r.h.s. is related with the social imitation process and
the second one with the personal differences in the evaluation about the
relative advantages of the adoption. As we show in the reference \cite%
{Laciana}, when the commercial target has decision makers with different
psychological characteristics, a weight factor multiplying each difference
individually of Eq. (\ref{9}) can be added.

As an example we consider the particular case of $T=0$ ($\beta =\infty $).

From Eq. (\ref{7}) the following decision algorithm results

\begin{equation}
P\left( \overrightarrow{S}_{\alpha }=\overrightarrow{X}_{k}\right) =\left\{ 
\begin{array}{c}
1\text{ if }\Delta _{kj}>0\text{ }\forall \text{ }j\neq k\text{ \ \ \ \ \ \
\ \ \ \ \ \ \ \ \ \ \ \ \ \ \ \ \ \ \ \ \ \ \ \ \ \ \ \ \ \ \ \ \ \ \ \ \ \
\ \ } \\ 
\frac{1}{1+l}\text{ if }\exists \text{ }l\text{ values of }j\text{ with }%
\Delta _{kj}=0\text{ and }\Delta _{kj}>0\text{ for all others} \\ 
0\text{ \ if \ }\Delta _{kj}<0\text{ for any }j\neq k\text{ \ \ \ \ \ \ \ \
\ \ \ \ \ \ \ \ \ \ \ \ \ \ \ \ \ \ \ \ \ \ \ \ \ \ \ \ \ }%
\end{array}%
\right.  \label{10}
\end{equation}

\subsection{Ising-like model as a particular case (M=2)}

In this case there are only two states, i. e. $E=\left\{ \left( 1,0\right)
,\left( 0,1\right) \right\} \equiv \left\{ \overrightarrow{X_{1}},%
\overrightarrow{X_{2}}\right\} .$ Then Eq. (\ref{9}) becomes:

\[
\Delta _{12}=\nu _{1}-\nu _{2}+u\left( \overrightarrow{X}_{1}\right)
-u\left( \overrightarrow{X}_{2}\right) . 
\]

In the simplest case of T=0, using Eq. (\ref{10}) we obtain

\begin{equation}
P\left( \overrightarrow{S}_{\alpha }=\overrightarrow{X}_{1}\right) =\left\{ 
\begin{array}{c}
1\text{ if }\Delta _{12}>0 \\ 
1/2\text{ if }\Delta _{12}=0\text{ \ \ } \\ 
0\text{ if }\Delta _{12}<0%
\end{array}%
\right.  \label{11}
\end{equation}

The algorithm given by Eq. (\ref{11}) has been used in ref.,\cite{Laciana}
for the calculation of the penetration of a given new product in a potential
market. The options in that case were only adoption or non-adoption.

\subsection{Decision algorithm for the studied cases (M=3 and M=4)}

In what follows, we will adapt Eq. (\ref{7}) to specific
examples,determining the state of each agent according to the probability
resulting from it.

\subsubsection{Case of three options}

We will apply the formalism developed before to the case of three options;
the adoption of the product 1, the adoption of the product 2 or the
non-adoption. We assume that both products are in competition for the same
potential market. Then the set of states will be given by

\begin{equation}
E\left\{ \left( 1,0,0\right) ;\left( 0,1,0\right) ;\left( 0,0,1\right)
\right\} \equiv \left\{ \overrightarrow{X_{1}};\overrightarrow{X_{2}};%
\overrightarrow{X_{3}}\right\} .  \label{12}
\end{equation}

The adoption probability of agent $\alpha $, for example, in relation with a
generic product 1 in the t instant, with a noise level in the decision given
by $\beta $, is obtained by the simple application of Eq. (\ref{7}),
obtaining the following expression

\begin{equation}
P\left( \overrightarrow{S}_{\alpha }\left( t\right) =\overrightarrow{X}%
_{1}\right) =\frac{1}{1+e^{-\beta \Delta _{12}}+e^{-\beta \Delta _{13}}}.
\label{13}
\end{equation}

\subsubsection{Case of four options}

We will also consider an application example with four options, in order to
illustrate the adaptability of this formalism to more complex problems. The
application is completely analogous to the previous case, except for the
additional state in the set:

\begin{equation}
E\left\{ \left( 1,0,0,0\right) ;\left( 0,1,0,0\right) ;\left( 0,0,1,0\right)
;\left( 0,0,0,1\right) \right\} \equiv \left\{ \overrightarrow{X_{1}};%
\overrightarrow{X_{2}};\overrightarrow{X_{3}};\overrightarrow{X_{4}}\right\}
.  \label{14}
\end{equation}

Where one of the states corresponds to non-adoption.\emph{\ }

The probability of agent $\alpha $ taking, for example, state $%
\overrightarrow{X_{1}}$, as deduced from Eq. (\ref{7}), will be:

\begin{equation}
P\left( \overrightarrow{S}_{\alpha }\left( t\right) =\overrightarrow{X}%
_{1}\right) =\frac{1}{1+e^{-\beta \Delta _{12}}+e^{-\beta \Delta
_{13}}+e^{-\beta \Delta _{14}}}.  \label{15}
\end{equation}

\section{Variables of the model}

The microscopic variables introduced in the diffusion model are related
essentially with: the way in which the agents are in communication (mainly
by means of the kind of social network), the individual evaluation about the
adoption of a new product (personal perception), the initial rate of
innovators (mainly influenced by campaigns of advertising) and finally the
introduction of an additional parameter (temperature) for the quantification
of the uncertainty in the decision process.

\subsection{Rewiring probability of the SWN}

We will use Watts and Strogatz's \cite{Watts Strogatz} method to generate a
family of networks, from the regular network to the one totally random,
passing by the set of networks known as small world networks (SWN). The SWN
has similar properties to the social real networks, such as is stressed in
ref. \cite{Albert}.

We begin with a regular lattice in two dimensions, with interaction between
the eight nearest neighbors for each agent (neighborhood of Moore), then we
consider the possibility of re-connection between an agent and another one
outside the neighborhood, introducing a re-wiring probability. This
methodology lets us study how the topological changes in the social network
influence the adoption process.

\subsection{Utility of products in competition}

We use the concept of utility in a broad sense as the value that the
decision-maker gives to the new product. The functional form of this
variable will depend on the type of considered article, i.e. by the
different attributes that characterize the product in question. For example
in the case of a car the attributes could be: price, efficiency, comfort,
safety, speed, etc. Each one of those attributes could be pondered in a
different way depending on the individual. Utility is a convenient concept
to describe the subjective personal decision. Another example, totally
different to the one given before, are financial assets. It is evident in
this case that decision is strongly induced by the risk aversion of each
individual \cite{Schoemaker} \cite{Pratt} \cite{Machina}.

The definition of a utility function is in general a complex task in the
field of the economy, for that reason this subject is not analyzed in the
present work. In our case we introduce a parameter which takes values
between 0 and 1 which plays the role of utility. In our approach the
utilities of the two products considered will be respect to the utility of
non-adoption.

\subsection{Spatial and temporal distribution of early adopters}

The technology diffusion model proposed has, for a given agent an implicit
adoption threshold, which will be related to the number of adopters within
its group of contacts (neighbors for a regular net) and to the personal
evaluation about the advantages of the adoption. For example, such it is
shown in ref. \cite{Laciana}, for the case of the introduction of only one
new product (M = 2), when the difference of utility is 0.6, in a regular
network with an interaction until 8 neighbors, the adoption threshold is of
two adopters in the neighborhood, while for a difference of utility of 0.8
it is 1. Then we can say that the change in the distribution of the initial
adopters in space modifies the possibilities to reach, for a given agent,
the adoption threshold, just as it is shown in ref. \cite{Weisbuch}, \cite%
{Laciana} and with it the time in that the product reaches the saturation of
the potential market.

The rate at which innovators are generated is also important, as it strongly
affects the takeoff time. The innovators will be introduced, in the present
work, in a linear way until we reach the 2.5 \%, in agreement with ref. \cite%
{Rogers}. That proceeding was introduced in ref. \cite{Laciana et al.}.

\subsection{Temperature as a measure of uncertainty in the decision}

\textquotedblleft Temperature" in the social system considered represents
the global uncertainty regarding the decision. With temperature different to
0, via Eq. (\ref{7}), the decision process about the adoption or not of the
new products (or technologies) becomes more stochastic. In fact that effect
can be interpreted as noise, due to erratic circumstances which have an
influence in the opinion of all agents, in the moment of performing the
decision \cite{Grabowski}, \cite{Schweitzer}. For example, if the agents are
agricultural producers deciding whether to adopt of a new seed, the
temperature could represent fluctuations due to epidemics, weather
variations, political events, etc. \cite{Weisbuch}. As consequence of those
events the decision is not taken with normality.

\section{Agent model implementation}

Initially we will consider a regular square grid of 200 X 200, i.e. 40,000
agents. We are not assuming periodic conditions in the boundaries.

Agents will be initially connected in a regular lattice, and through Watts
and Strogatz's \cite{Watts Strogatz} method of re-wiring we will replicate
small world networks. For the implementation of the model we will use the
software known as Anylogic \cite{Anylogic}.

In all the cases the decision algorithm is coming from Eq. (\ref{7}).

\section{Results}

By means of numerical experiments we will study how the penetration, in a
common market, of products in competition is influenced by the variation of
the following microscopic variables: the probability of rewiring $P_{r}$,
the difference of utilities between adopting and not adopting $\Delta u$,
the initial rate of innovators $\gamma $ and the temperature $T$ as a
measure of the uncertainty in the decision process.

Also we will analyze the effects, in terms of market penetration, of
improving a product, but delaying the launch in relation to another product
with no improvement, but that it is launched without delay.

Finally, in order to explicitly show the versatility of the proposed
formalism, we will make an experiment with four options. In this case, we
will consider two not mutually exclusive products (adopting both of them is
a possibility) in a common market, and analyze the influence of variables $T$
and $P_{r}$ in the diffusion curves.

\subsection{Experiments with three options}

We will consider the introduction of two products in a common potential
market. Therefore in our dynamical system we have three kinds of agents;
those that adopt product A, those that adopt product B, and those that do
not adopt any of both products. We can call the last option 0. In our
proposal the dis-adoption is not included, the following transitions are not
allowed: A $\rightarrow $ B, B $\rightarrow $ A, A $\rightarrow $ 0 and B $%
\rightarrow $ 0. The allowed transitions are: 0 $\rightarrow $ A and 0 $%
\rightarrow $ B.

The probability of occurrence of those transitions is given by Eq. (\ref{13}%
), with the states given by $\overrightarrow{X}_{1}$, $\overrightarrow{X}%
_{2} $, $\overrightarrow{X}_{3}$ corresponding to the options A, B and 0
respectively.

\subsubsection{Influence of $P_{r}$ in the adoption pattern}

In this numerical experiment we will consider two very different values for
the probability of rewiring, one corresponding to the regular network ($%
P_{r} $ = 0.) and another above the superior limit of the SWN ($P_{r}$ =
0.02). We will also suppose that the evaluation that the agents perform
about the advantages of adopting one or another product is the same for both
products and for all agents (homogeneous market) i.e. $\Delta u_{A}$= $%
\Delta u_{B}$= 0.6 for all the agents. We will also suppose that the
innovators are introduced to complete the 2.5\% of the market with a
temporal rate ($\gamma $) of 125 by tick. The results are shown in Fig. 1 a)
and b):

\begin{figure}[ptb]\begin{center}
\includegraphics[
natheight=8.1872in, natwidth=8.0004in, height=2.8781in, width=2.8132in]
{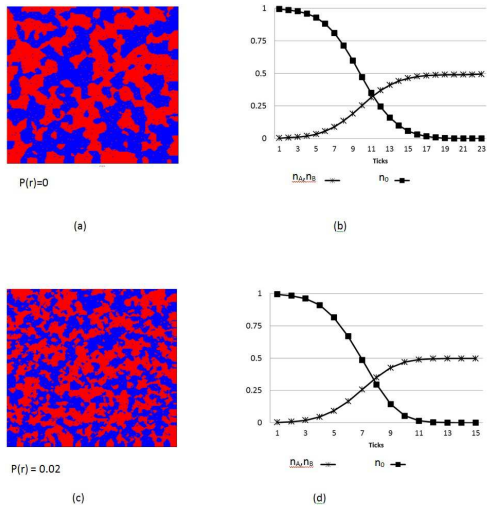}%
\caption{Adoption patterns and graphics for the proportion of adoption of
product A (n$_{A}$), proportion of adoption of product B (n$_{B}$) and
proportion of non-adopters (n$_{0}$) as a funcion of time (ticks). In (a)
landscape of adoption at saturation, with rewiring probability $P_{r}=0,$in
(b) curves $n_{A}$, $n_{B}$ and $n_{0}$ vs t with $P_{r}=0,$ in (c) adoption
landscape for $P_{r}=0.02$, and in (d) $n_{A}$, $n_{B}$ and $n_{0}$ vs t
when $P_{r}=0.02.$}
\end{center}\end{figure}%

This simulation serves firstly as a way of testing the model, since as the
considered products which have the same utility and time of launching, they
should present the same adoption curve. That in fact occurs as can be
observed in Fig. 1 a,b. Secondly, the adoption distribution pattern in the
grid at the end of the adoption process, as can be seen in the figure, is
more homogeneous in the case of biggest rewiring probability. Finally, we
can observe that the saturation time is smaller in the case of biggest
rewiring probability. This behavior corresponds with the fact that the path
length and the clustering coefficient of the network are smaller for the
biggest rewiring probability \cite{Watts Strogatz}.

\subsubsection{Dependence of the adopter proportion with the innovators
generation rate}

Following the methodology introduced in ref. \cite{Laciana et al.} the
innovators are introduced according to Roger \cite{Rogers} until reaching
2.5\% of the market. We want to see how the way of introducing the
innovators affects the proportion of the market reached at the end of the
process. In order to study that effect we employ a different incorporation
rate of innovators for each product. Product A is introduced at a rate of
125 by tick (i.e. $\gamma $ = 125), this way 8 ticks are necessary to
complete the total number of innovators, while for the product B we
considered values of $\gamma $ from 125 until 1000 (where all innovators are
all introduced in the first tick). From an operative point of view, this
could mean an aggressive advertising campaign of product B before its
launching. The results are shown in the Fig. 2.

\begin{figure}[ptb]\begin{center}
\includegraphics[
natheight=7.427in, natwidth=8.5106in, height=2.8772in, width=3.2932in]
{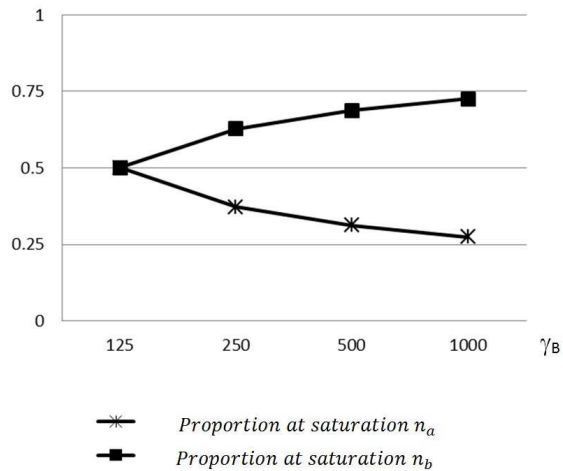}%
\caption{$n_{A}$ and $n_{B}$ at saturation, vs the rate of innovators of
product B. }
\end{center}\end{figure}%

We can see from Fig. 2 that the influence of the microscopic variable $%
\gamma $ is decisive in the possibility of obtaining a bigger portion of the
market. Product B almost reaches 75\% of the market although the only
difference between the products A and B would be, for example, a different
advertising campaign in the launching which enables product B to begin the
competition with a bigger number of innovators. In the studied example both
products have the same difference of utility (0.6) in relation to no
adoption. We have repeated the numerical experiment with the variation of
other quantities such as the temperature and the rewiring probability
without observing considerable differences in the final result. We therefore
conclude that the most important effect is the one associated to the
variable $\gamma $.

\subsubsection{Tradeoff between improvement and time of launching}

We will consider two products, A and B, competing for the same potential
market. A is launched at time $t=0$, while B is launched at a posterior time 
$t=t_{B}$, but, during that time difference, B is improved. In reference 
\cite{Bhattacharya} two practical possibilities for the use of the extra
time previous to the launching are presented, one would be to decrease the
unitary cost and the other is our case, where the time is employed to
improve the product.

Let us call the difference of utility between adopting product A or not $%
\Delta u_{A}$ and the analogous for product B; $\Delta u_{B}.$ During time $%
t_{B}$ product B is improved and we assume that the utility $\Delta u_{B}$
is an increasing function of $t_{B}$, with $\Delta u_{B}$($t_{B}=0$) = $%
\Delta u_{A}$ as an initial condition. Then we propose the following
expression for the utility:

\begin{equation}
\Delta u_{B}=\Delta u_{A}+\left( 1-\Delta u_{A}\right) \tanh \left( \frac{%
t_{B}}{\tau }\right) .  \label{16}
\end{equation}

Parameter $\tau $ gauges the difficulty of improving the product, and has
units of time. As we see from Eq. (\ref{16}), when $t_{B}$ $\rightarrow
\infty $ the utility $\Delta u_{B}\rightarrow 1$ and when $t_{B}$ $%
\rightarrow 0$ the utility $\Delta u_{B}\rightarrow \Delta u_{A}$. We assume
that the utility is normalized such that $0\leqslant \Delta u\leqslant 1.$

Numeric experiments will be performed considering $\Delta u_{A}=0.6$, $%
P_{r}=[0,0.02]$ (the second value of rewiring probability corresponds to the
maximum value of the SWN) and $T=[0,0.05]$ for the temperature. The trials
will be separated in two subsections corresponding to the following cases:
a) when the population is homogeneous, in relation to the evaluation of the
utility, and b) when the population is heterogeneous. We assign the
convenient value of 20/3 to $\tau $.

\paragraph{Homogeneous population of decision makers}

In this case all the decision makers evaluate both products in the same way.
Then, for all the agents, we assume that product A has a utility $\Delta
u_{A}=0.6$ and that the advantages of product B are increased respect to A
according to Eq. (\ref{16}), starting from the value 0.6. In this sense, the
potential market can be considered as a homogeneous population. Under this
supposition the graphics of the Fig. 3 were obtained.

\begin{figure}[ptb]\begin{center}
\includegraphics[
natheight=8.1872in, natwidth=8.0004in, height=2.8781in, width=2.8132in]
{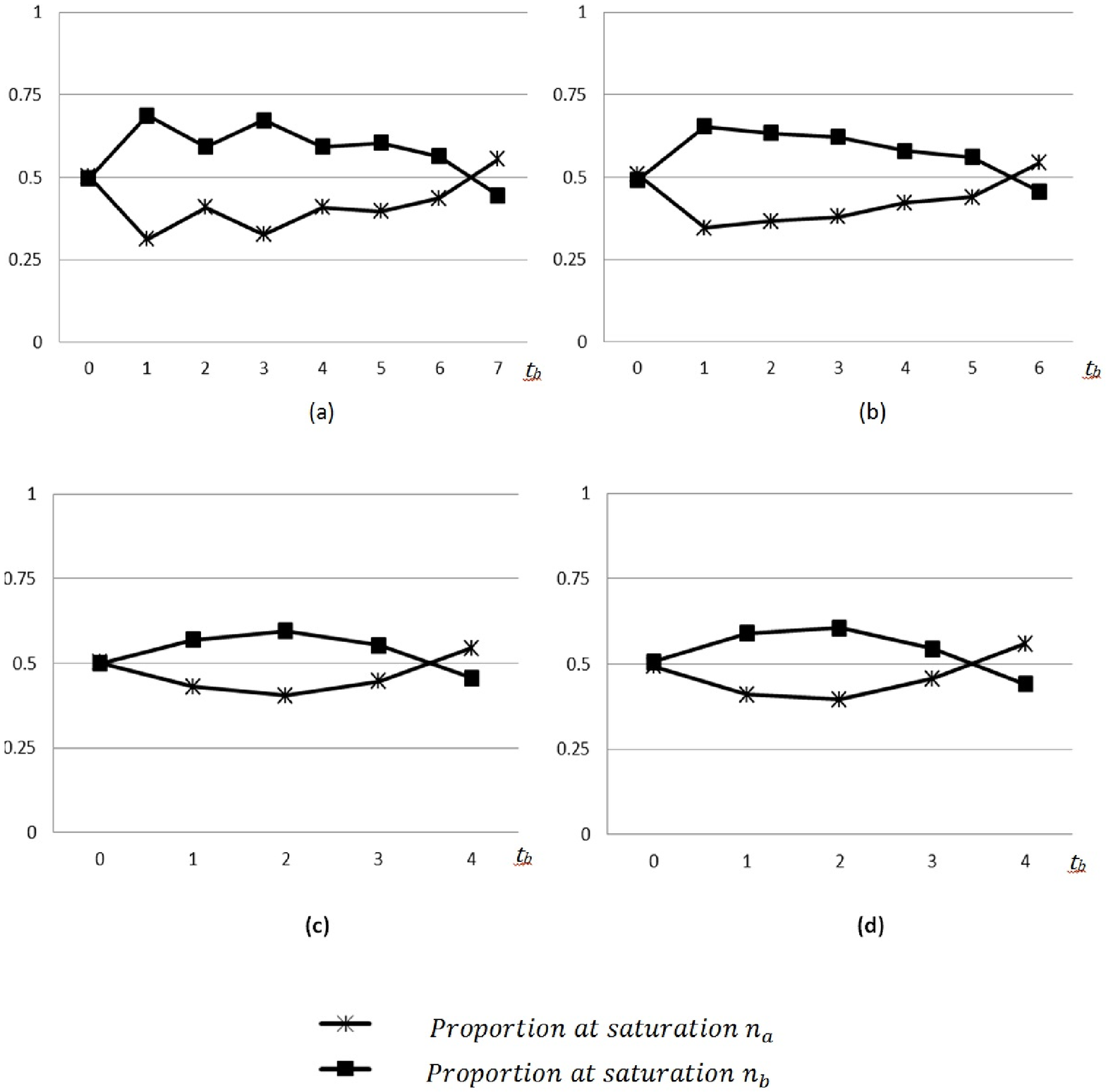}%
\caption{$n_{A\text{ }}$and $n_{B\text{ }}$at saturation for different
launching times of B, for a homogeneus set of decision makers. In (a) $%
P_{r}=0$ and $T=0$, in (b) $P_{r}=0.02$ and $T=0$, in (c) $P_{r}=0$ and $%
T=0.05$, and in (d) $P_{r}=0.02$ and $T=0.05.$}
\end{center}\end{figure}%

In Fig.3a, that corresponds to $P_{r}=0$ and $T=0$, we observe a
non-realistic pattern. The maximum difference of market proportion between
the products is obtained at $t_{B}$=1. At $t_{B}$=2 the difference
diminishes, but it increases at $t_{B}$=3 again. That occurs because the
adoption threshold changes in $t_{B}$=1 in the same way for all the agents
and it is necessary to wait until $t_{B}$=3 to observe a new threshold
change. Therefore, the delay in launching between $t_{B}$=1 and $t_{B}$=2
has no positive effect on the diffusion of product B at all, allowing
product A to gain extra market share. That occurs because the decision
makers, under the homogeneity assumption, act all in the same way; this
behavior would be not realistic.

As will be seen in the next subsection, when a heterogeneous population of
deciders is considered, the pattern mentioned before is not observed.
Another way of introducing heterogeneity and thus avoiding this
non-realistic pattern, is to give values different than zero to the rewiring
probability Pr or to the temperature, as shown in figures 3 b), 3 c) and 3
d).

When rewiring is performed, the average number of connections per agent
remains constant, which in our case is of 8 nearest neighbors. However, at
an individual level, this number is not maintained. This fact generates
heterogeneity in the population, which is manifested through the individual
network of contacts. Therefore, when product B is improved, each agent's
answer is unique, contrary to the `in block' response of a homogeneous
population of decision makers, i.e. some of them reach the following
threshold later than in the regular lattice compensated by others that reach
it earlier. The result of that compensation can be seen on the Fig. 3 b),
where the curve flattens out after the first tick compared to Fig. 3 a).
Another thing that we can observe in Fig. 3 b) comparatively with Fig. 3 a)
is that the critical launch time, time limit below which product B obtains
most of the market, is reached one tick before than in the case of 3 a).

We emphasized before that the temperature adds uncertainty to the decision,
this means that reaching the threshold doesn't assure that the adoption is
produced and not reaching the threshold doesn't invalidate the possibility
of adopting. As a consequence there is a tendency of reducing the
differences between the adopter proportion of A and B. \ As we can see in
Fig. 3 c) the graph is more flattened, the maximum difference of proportions
is now for $t_{B}=2$ and the critical point, where inversion of the
population proportions occurs, is closer to $t_{B}=3.$ Finally, in Fig. 3 d)
the combined effect of $P_{r}\neq 0$ and $T\neq 0$ is observed. The only
difference with Fig. 3 c) is that the maximum difference of proportions
appears at $t_{B}=1.$

\paragraph{Heterogeneus population of decision makers}

As we mentioned before, when a homogeneous population of potential adopters
is considered, the patterns obtained are not very realistic. In this
subsection we assume that agents have individual perceptions of the utility
of product A, i.e. $\Delta u_{A},$ has the average value of \TEXTsymbol{<}$%
\Delta u_{A}$\TEXTsymbol{>} = 0.6 , but at an individual level there are
some differences given by the following distribution: 40\% of the deciders
with $\Delta u_{A}$= 0.6, another 40\% with $\Delta u_{A}$= 0.7 and the
other 20\% with $\Delta u_{A}$= 0.4. For product B, the distribution is the
same than A, when the delay in the launching is not taken into account, and
when it is, the distribution is modified in agreement with Eq. (\ref{16}).

The same numerical experiments that in the homogeneous case were performed
and the results are shown in the Fig. 4.

\begin{figure}[ptb]\begin{center}
\includegraphics[
natheight=8.6351in, natwidth=8.0004in, height=2.8772in, width=2.6688in]
{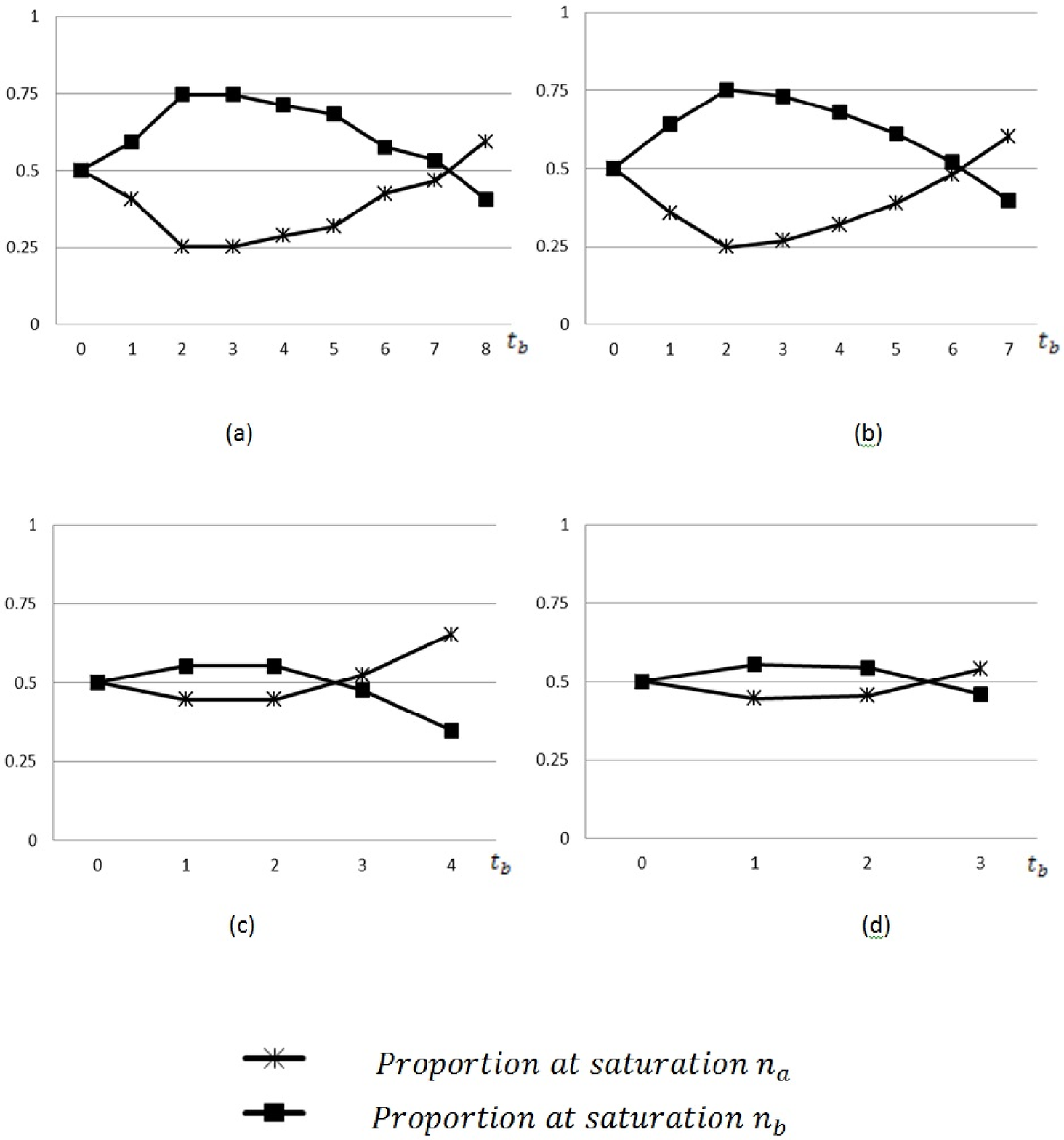}%
\caption{$n_{A\text{ }}$and $n_{B\text{ }}$at saturation for different
launching times of B, for a heterogeneus set of decision makers. In (a) $%
P_{r}=0$ and $T=0$, in (b) $P_{r}=0.02$ and $T=0$, in (c) $P_{r}=0$ and $%
T=0.05$, and in (d) $P_{r}=0.02$ and $T=0.05.$}
\end{center}\end{figure}%

In Fig. 4 a) we can see that the biggest difference between both populations
is produced when the delay in the launching of B is of two ticks, and that
the population of B is bigger than that of A until the seven ticks are
reached. After that, the relationship between populations is reverted.

In the Fig. 4 a) we can observe the existence,\ at $t_{B}=2$, of an absolute
maximum, contrary to what happens with Fig. 3a), where two local maxima
exist. This is not modified in the other experiments: in Fig. 4 b) for
example, corresponding to $P_{r}\neq 0$, the maximum difference of
populations between both products happens for a delay of two ticks in the
launching of B. The only change is the decrease, in approximately a tick, of
the time for the population of the adopters of product A to overcome the
population of adopters of product B. This would be a critical point beyond
which the launching is no longer convenient.

The variable that produces a drastic change is temperature, as can be
observed in figures 4 c) and 4 d). The uncertainty in the decision,
quantified by the temperature, distorts the relative advantages of a product
compared to the other. This way of introducing uncertainty is the simplest.
A more realistic form would be to keep in mind each agent's differences
regarding the risk. However, for simplicity, in this analysis we will
suppose that all the agents are affected in the same way. In such sense,
temperature can be thought of as a global parameter, like inflation or
dollar quotation. The process of decision will become then more stochastic
than automatic and causal.

In graph 4 c) we see that the improvement of product B does not assure a
great advantage for reaching a larger part of the market. The maximum
difference is between $t_{B}$=1 and $t_{B}$=2 , being approximately of only
10\%, while in the example of null temperature it was around 50\%, as we can
see in graphs 4 a) and b). Moreover, when $T\neq 0$ a delay of 3 or more
ticks in the launching is an unfavorable strategy.

In Figures 3 and 4 we can see that improving the product has a positive
effect in its market penetration until a \textquotedblleft critical
time\textquotedblright , where the further delay of the improved product
would make it give up the market leadership.

By comparing figures 3 a) and 3 b), we can see that increasing the
randomness of the network (\textit{P}$_{r}$\textit{\ = }0\textit{\ }$%
\rightarrow $\textit{\ }0.02) reduces the critical time. This behavior is
observed with the same intensity even in the heterogeneous case, as we can
see in figures 4 a) and 4 b), in this case the critical time being one unit
larger.

Adding uncertainty to the decision through temperature, strongly affects the
value of the critical point. Comparing Fig.3 a) with 3 c) we can observe how
this point decreases by three units. The same effect is seen when
considering a more realistic social network (\textit{P}$_{r}$ = 0.02) in
figures 3 b) and 3 c). These behaviors, associated to the change of
temperature, show us how a more uncertain scenario adds risk to the
launching strategy, making the choice of further development disadvantageous.%
\emph{\ }

It should be noted that in an uncertain scenario ($T$ \TEXTsymbol{>}0) the
type of network becomes less important as can be seen in figures 3 c) and 3
d), where the critical time is not altered.

The observed effects in uncertain scenarios becomes more dramatic for a more
realistic model with the possibility of a heterogeneous population, as can
be seen by comparing figures 4 a) with 4 c), and 4 b) with 4 d). We can also
conclude in this case that the effect of changes on the network topology is
negligible when uncertainty is considered.

\subsection{Experiments with four options}

As an example that shows the versatility of the formalism we propose, we
will analyze a case in which the decision-makers choose between four
options. These options will be of not mutually exclusive durable goods, such
that agents can choose product A, product B, both A and B or non-adoption.
We could imagine, as a possible example, that agents are household heads
that must decide between buying a mid-range car (A), a high-end car (B) or
both. If we are speaking of middle class households, the cost of the
purchase and of fixed costs would be determining attributes in the decision
process.

Our focus will be in analyzing how the network topology (by varying
parameter \textit{P}$_{r}$) and the uncertainty of the socio-economic
scenario (by varying parameter $T$) affect the final pattern of adoption.

As a simplification we will assume that decision-makers evaluate the options
in the same way (we could think of average agents). We will then have all
potential consumers that consider a difference of utility\emph{\ }$\Delta
u_{A,0}$ between adopting product A and non-adoption (0), a\emph{\ }$\Delta
u_{B,0}$\emph{\ }for product B, and a\emph{\ }$\Delta u_{AB,0}$\emph{\ }for
the adoption of both products. For a more realistic approach, instead of
considering average decision-makers, using distributions of utility that
take into account social classes should be used.

The value of the utilities in our case will be assigned arbitrarily, as this
is simply an academic example, but in a real problem, a fine analysis of the
attributes of each product would be required. The only assumption respecting
the chosen values is that, for this specific group of households, the
high-end product will be of a lower utility that the mid-range, and that the
acquisition of both will have the lower utility. This order should be
modified if we considered high class decision-makers.

We will also assume that, considering that A and B are durable goods,
disadoption is not allowed, but as A and B are not mutually exclusive, A$%
\rightarrow $AB and B$\rightarrow $AB are admitted transitions. We will also
consider the 0$\rightarrow $AB transition, that is, the switch from a state
of non-adoption to the acquisition of both products simultaneously.

Therefore, in the decision algorithm given by\emph{\ }Eq.\emph{\ }(\ref{15})%
\emph{, }intervene, through\emph{\ }$\Delta _{jk}$, the differences of
utility\emph{\ }$\Delta u_{A,0}$, $\Delta u_{B,0}$, and $\Delta u_{AB,0}$,
associated to transitions\emph{\ }$0\rightarrow A$, $0\rightarrow B$, and%
\emph{\ }$0\rightarrow AB,$ and\emph{\ }$\Delta u_{AB,A}$, and\emph{\ }$%
\Delta u_{AB,B}$ corresponding to \emph{\ }$A\rightarrow AB$ \emph{y }$%
B\rightarrow AB$ respectively. These latter differences are calculated using
the first, since:

\begin{equation}
\Delta u_{AB,A}=u_{AB}-u_{A}=\Delta u_{AB,0}-\Delta u_{A,0}.  \label{17}
\end{equation}

\begin{equation}
\Delta u_{AB,B}=u_{AB}-u_{B}=\Delta u_{AB,0}-\Delta u_{B,0}.  \label{18}
\end{equation}

In our example, taking into account the considerations previously mentioned
of the order of preference of each of the options, we assign\emph{\ }$\Delta
u_{A,0}=0.7$, $\Delta u_{B,0}=0.65$, and\emph{\ }$\Delta u_{AB,0}=0.6$.

We will perform four experiments, with \textit{P}$_{r}$ values of 0 and 0.02
(regular lattice and small world network) and $T$ values of 0 and 0.05
(deterministic scenario and moderate uncertainty scenario).

The results are shown in Figures 5 a,b,c and d.

\begin{figure}[ptb]\begin{center}
\includegraphics[
natheight=7.427in, natwidth=8.5106in, height=2.8772in, width=3.2932in]
{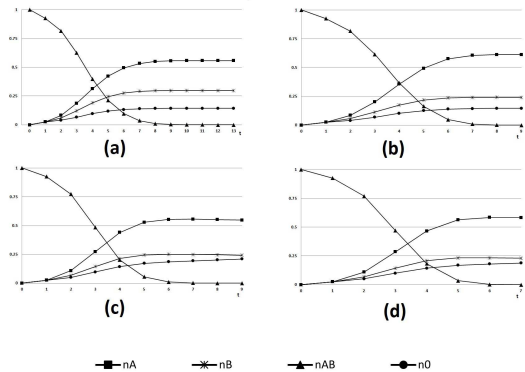}%
\caption{Diffusion proportion vs time; a) \textit{P}$_{r}$=0., $T$=0, b) 
\textit{P}$_{r}$=0.02, $T$=0, c) \textit{P}$_{r}$=0, $T$=0.05, \textit{P}$%
_{r}$=0.02, $T$=0,05.}
\end{center}\end{figure}%

By observing figures 5a and 5b we can see that increasing \textit{P}$_{r}$
has a positive effect on the diffusion of product A. This seems reasonable,
as it would show that a more efficient network increases the imitation
effect, and therefore favors the more massive choice, in our case, the
mid-range, more accessible car. This, in turn, affects the diffusion of the
high-end car (product B). The number of buyers of both products stays
virtually unchanged.

When uncertainty is added to the model (Fig. 5 c)), we observe that some of
the consumers of product B decide to adopt A as well. This could infer that
the increased uncertainty leads decision-makers to a less-rational behavior,
making them choose an operation with higher risk involved.

Finally, in Fig. 5 d) we see how an increase in \textit{P}$_{r}$ and $T$
simultaneously, produces a compensation of effects in the proportion of
adopters.

An important result of these experiments is related to the differences in
times of market saturation for each of the scenarios, which can be evaluated
approximately by analyzing the non-adoption curves when varying \textit{P}$%
_{r}$ and $T$. We can see that increasing either of them decreases the time
of saturation strongly. In the case of the rewiring probability, the more
efficient communication between agents favors the imitation effect, and thus
the speed of diffusion. In the case of temperature, some transitions that
have null probability of occurrence when $T$ = 0 become slightly possible
with $T$ \TEXTsymbol{>} 0, and this reduces the diffusion time sharply.

\section{Conclusions}

A formalism has been developed that allows the study of systems formed by
individuals that must decide among several options. This formalism is
sufficiently versatile for it to include heterogeneous populations of
decision makers deciding between many different options. In our work, that
methodology is applied to a problem of three options; the adoption of a
product A, a product B, or the non-adoption. It is also applied to a four
options problem where the possibilities are: adoption of product A, of
product B, of both, or no adoption.

From the numerical experiments performed, for three options case, we can
extract the following conclusions:

\begin{description}
\item[*] As the social network becomes more random, the market becomes
saturated with product buyers more quickly.

\item[*] A quicker generation of innovators, for example through a bigger
investment in publicity, modifies the curve of adoption of the product
drastically, reaching as a consequence, in the best of cases more than 70\%
of the market. The final proportion reached, is slightly affected by the
uncertainty in the decision (which is introduced through a parameter
analogous to the temperature of the statistical system). The final
proportion practically does not change with the modification of the topology
of the social network (by means of the rewiring probability). However these
factors accelerate the arrival to the equilibrium.
\end{description}

We have also performed numerical experiments related with the diffusion of
two products in the same potential market, but where one of the products is
launched with a delay, and during this time this product is being improved.
We analyze then, when the market is saturated, the advantage obtained by the
improved product. In relation to these experiments we conclude the following:

\begin{description}
\item[*] When a homogeneous set of agents is considered, in relation to
their evaluation of the comparative advantages of the products, unrealistic
results are obtained, due to the emergence of a collective threshold of
decision.

\item[*] Temperature, or the inverse of the confidence coefficient \cite%
{Bourgine}, is the variable that produces the biggest effect when changed.
This variable is associated with the noise or uncertainty in the process of
decision, which can be related with certain socio-economic scenarios of high
volatility. From the results we can see that temperature causes an
appreciable reduction of the advantages of product improvement, favoring the
one that was launched first. It also reduces the interval of delay in
launching ($t_{B}$) for which an advantage in the won market proportion is
obtained at market saturation.
\end{description}

A topic to investigate in the future, by means of the application of the
developed methodology, would be for example, the resulting effect of the
investment in advertising (via the innovators generation) in scenarios with
certain degree of uncertainty and distributions of decision makers with
different risk aversion. This is only one of many possible cases of
numerical experiments that are facilitated, without big computing efforts,
by agent based modeling.

In regards to the four options problem, we observe that:

\begin{description}
\item[*] With a larger rewiring probability, the adoption of the product
with greater utility is favored. In other words, the reduction in the mean
characteristic path length of the network increases the imitation effect,
and leads to a massification of the most convenient product.

\item[*] An increase in uncertainty (temperature greater than 0) increases
the probability of adopting the combination of both products.
\end{description}

The patterns we observe using this model show no unexpected behaviors,
however, future investigations should compare these results with real
processes, for validation purposes.

\section{Acknowledgments}

This investigation has been enriched by comments and suggestions of an
anonymous revisor. This research was supported partially by two U.S.
National Science Foundation (NSF) Coupled Natural and Human Systems grants
(0410348 and 0709681) and by the University of Buenos Aires (UBACyT 00080).

\end{document}